\documentclass[runningheads]{llncs}

\usepackage[T1]{fontenc}
\usepackage{graphicx}
\usepackage{color}

\usepackage{graphicx}
\usepackage[caption=false]{subfig}

\usepackage{tabularx}
\usepackage{multirow}
\usepackage{booktabs}

\begin{document}

\title{Raise Awareness of the Environmental Impacts\\of Retail Food Products: A User-Centered Scenario-Based Approach}
\titlerunning{Raise Awareness of the Environmental Impacts of Retail Food Products}

\author{Lorenzo Porcelli%
\and
Francesco Palmieri%
}
\authorrunning{L. Porcelli and F. Palmieri}
\institute{Department of Computer Science, University of Salerno, Fisciano (SA), Italy\\
\email{\{lporcelli, fpalmieri\}@unisa.it}}
\maketitle             
\begin{abstract}
The climate is warming rapidly, and atmospheric concentrations of greenhouse gases (GHGs) are at their highest levels ever recorded. As a result of these climate changes, caused mainly by human activities, disasters have increased fivefold over the past 50 years, causing death and economic loss. Civic engagement and awareness are essential to mitigate climate change and its impacts. In this work, we proposed a user interface that makes users aware of the environmental impact of the food products they buy when shopping. A user-centered scenario-based design was followed in the development of the interface. Gamification elements were added to increase civic participation in climate action.

\end{abstract}

\section{Introduction}
Climate change is one of the most pressing challenges that humanity is currently facing. The last eight years were the warmest on record \cite{warmestyears}. The Sustainable Development Goals Report 2022 \cite{united2022sustainable} estimates that 700 million people will be displaced by drought alone by 2030, and about one-third of the world's land area will experience at least moderate drought by 2100. The most significant impact will be poverty and hunger, as millions of people won't have access to fundamental services such as healthcare and education. As well as hindering economic growth, increasing inequality can be a source of international conflicts.

To address the world's most pressing challenges, the United Nations proposed a global call for sustainable action to improve people's lives and preserve the planet identifying 17 Sustainable Development Goals (SDGs)\footnote{Sustainable Development Goals (SDGs): \url{www.un.org/sustainabledevelopment/}}. Among the SDGs, Goal 13 (Climate Action) focuses on actions to combat climate change and its effects. Climate change is, in part, a result of human activities that release greenhouse gases (GHGs). The GHGs are the main contributor to global warming. The primary objective of Climate Action is to strive for attaining net zero global greenhouse gas emissions by the year 2050. 

Carbon footprint is a term used to describe the total amount of greenhouse gases produced. The most common greenhouse gas is carbon dioxide. Carbon footprint also includes other gases such as methane, nitrous oxide and fluorinated gases.
By understanding and measuring our carbon footprint, we can identify areas where we can reduce greenhouse gas emissions and hence mitigate the impact on climate change.

Current national emission reduction commitments are insufficient to meet the targets set by the SDGs.
An important part of building the public support needed for successful climate change policy is raising public awareness \cite{drummond2018public}.
As at least 25\% of greenhouse gas emissions come from food \cite{owidenvironmentalimpactsoffood}, understanding the environmental footprint of retail food products can help consumers make more informed choices and reduce their impact. As consumers feed the food supply system, if they make informed choices and prioritize climate improvement, food producers need to adapt to consumer demands.

We proposed a graphical user interface for an application that informs consumers about the food product footprint they buy in a supermarket. In developing the interface we followed the scenario-based approach of Rosson and Carroll \cite{rosson2002scenario}. Gamification elements \cite{pelling2011short} were also added to increase user engagement.

The remainder of the paper is organized as follows. A review of related literature is presented in the next section. Section \ref{sec:food_footprint} provides a comprehensive definition of food product footprints and outlines several methodologies employed for their computation. Section \ref{sec:methods} describes the methodology followed to create a prototype graphical user interface for a mobile application. Finally, Section \ref{sec:conclusions} encompasses the conclusions drawn from the study and outlines potential avenues for future development.

\section{Related Work}
\label{sec:related_work}
A recent review \cite{rondoni2021consumers} of works that studied the impact of footprint labels on consumers showed that consumers have limited awareness of carbon-related measurements, and the current carbon footprint label system remains ambiguous. However, when redesigned using user-friendly symbols such as traffic light colors, consumers' understanding improves significantly.

However, the future of footprint labels is increasingly dematerialized. CarbonCloud\footnote{CarbonCloud: \url{https://carboncloud.com}} provides a free online database called ClimateHub that allows users to search and find information about the carbon footprint of 10,000 branded food and beverage products available in American grocery stores. Each product listed on their website includes a total emissions tag based on its weight. Consumers can access more detailed information by clicking on the product to learn about the percentage of emissions that come from transportation, packaging, processing, and agricultural practices.

It can sometimes be frustrating to check the footprints of purchased products one by one before shopping. Other proposals increase consumer awareness more quickly at the end of shopping.
Evocco\footnote{Evocco: \url{https://linktr.ee/evocco}} is a startup that has created a mobile app with the same name that helps users determine the carbon footprint of their food purchases and track the environmental impact of their choices. Users can take a photo of their grocery receipt, and the app's machine learning technology identifies the products and calculates their climate impact based on type, weight, and origin.

Our proposal aims to raise user awareness in the most direct way possible. We provide a tool as simple as possible and usable by anyone, which shows the footprint of a food product at the time that matters, i.e., when the product is about to be placed in the shopping basket.

\section{Food Product Footprints}
\label{sec:food_footprint}
Consumers are rarely shown the environmental consequences of producing and consuming food. The environmental footprint of product items allows consumers to compare the impact on the environment between different product groups.

The comprehensive environmental impact of producing a food product must take into account at least sustainability metrics such as carbon, nitrogen and water footprints \cite{leach2016environmental}. The carbon footprint of a product or service reflects the amount of greenhouse gas emissions released throughout its lifecycle, typically including production, use or consumption, and disposal \cite{roos2014carbon}. The nitrogen footprint of a food product indicates the overall quantity of reactive nitrogen released into the environment as a result of the production and consumption of that specific food product \cite{leach2012nitrogen}. The water footprint of a food product can be defined as the quantity of water used, both through evaporation and transpiration, during the production process of that particular product \cite{hoekstra2012global}. The work \cite{leach2016environmental} presents three calculation methods for food product footprints. 

The first calculation method, known as the footprint weight, directly shows the weight of the environmental impact of a specific product. Footprint weight is the most commonly used method when determining the environmental impact of a product. It is calculated as
\begin{equation}
    F_w = w \times f,
\end{equation}
where $F_w$ is the footprint calculation based on weight, $w$ is the weight of the product, and $f$ is the footprint factor.

The second calculation method measures the sustainability of a product based on the production process. It looks at specific sustainability measures, such as crop rotation, riparian buffers, and rotational grazing. When a producer meets more of these measures, the sustainability rating of the product increases. In the following formula, the percent of possible sustainability measures $F_s$ for a product is the sum of sustainability measures that apply at a given farm divided by the total possible sustainability measures, i.e.,
\begin{equation}
    F_s = \frac{\sum \textit{sustainability measures at farm}}{\sum \textit{all possible sustainability measures}}.
\end{equation}

The third method of calculating footprints, called \% Daily Value (DV), shows the percentage of a person's total daily footprint attributed to the consumption of a particular product. The \% DV is determined based on a reference value representing a sustainable daily environmental impact. The daily allotment of a healthy diet is another way of looking at the footprint of a single food product. The footprint as \% DV value can be written as
\begin{equation}
    F_{DV} = \frac{F_w}{D_w},
\end{equation}
where $F_w$ is the footprint of a product by weight, and $D_w$ is the total daily footprint associated with a healthy diet.

Each of these three footprint calculation methods can be represented graphically. A simple method proposed by \cite{leach2016environmental} is a star rating system that combines the three footprints into a single sustainability measure as an average of the three. The resulting star rating ranges from 0 stars (indicating the least sustainable) to 3 stars (indicating the most sustainable).

\section{Materials and Methods}
\label{sec:methods}
Our main objective was developing an effective interface for an app to increase awareness and civic participation in sustainability while shopping in available markets. We used a user-centric design paradigm to study the potential user base and consider it in all the conceptual design activities. A scenario-based process allowed us identifying the main actors involved and analyze their behavior. 

\subsection{Consumer behavior when shopping}
Recent work by \cite{machin2020habitual}, which collected data on 144 participants via eye tracking, confirmed the findings of previous studies regarding consumer behavior. Most consumers do not compare available products before buying but purchase the usual product without giving it much thought. Specifically, the average time consumers hold a product in their hands is less than the time needed to read the product information label. For those who compare products before purchasing, price stands out as one of the most important considerations.
Based on consumer behavior, we defined two personas described in Table~\ref{tab:personas}.
\begin{table}[ht]
\small
\renewcommand{\arraystretch}{1.5}
\setlength{\tabcolsep}{10pt}
\centering
\caption{Personas}\label{tab:personas}
\begin{tabular}{p{0.99\textwidth}}
\hline
(1) Maria is a 45-year-old housewife with a high school diploma. She goes shopping 2-3 times a week, usually in the morning between errands while her children are at school. As a mother of two children aged 8 and 10, she juggles many responsibilities, including managing the household and her children's schedules. Maria mainly chooses products by brand and convenience. She occasionally jots some things down to buy on a piece of paper, but most of her shopping is habitual. She primarily uses the smartphone to talk to her husband and children.\\
\hline
(2) Olivia is a 30-year-old woman with a degree in economics working as an accountant in a large company. She usually prefers shopping alone, either in the evening or at weekends. As a single professional, she has a busy schedule and values her free time. Olivia is tech-savvy and uses her mobile phone for both work and leisure. She is health conscious and buys organic and local produce whenever possible. She is organized and prepares her shopping list in advance using a checklist on her phone.\\
\hline
\end{tabular}
\end{table}

\subsection{Scenarios of current practices}
We hypothesized two scenarios, described in Table~\ref{tab:current_scenarios}, for defining the basic requirements and designing a potential solution. 

\begin{table}[ht]
\small
\renewcommand{\arraystretch}{1.5}
\setlength{\tabcolsep}{10pt}
\caption{Scenarios of current practices}\label{tab:current_scenarios}
\begin{tabular}{p{0.99\textwidth}}
\hline
(1) Maria gets up early and gets her children ready for school. She drops them off and goes shopping before picking them up in a few hours. As she enters, she is bombarded with several choices and options, but she has little time to waste. She navigates the store hastily, taking items from the shelves. She wants to save money and make sure she can get everything on her list before she goes to pick up her children from school. Having found everything, she goes to the checkout and leaves the store.\\
\hline
(2) It is Saturday morning, and Olivia checks her pantry to see what she needs to buy. She prepares her shopping list using a checklist app and heads toward the supermarket. On arrival at the supermarket, Olivia skims the aisles, comparing prices and reading labels to ensure the products meet her standards. As a health-conscious person, Olivia prefers organic products and checks the country of origin. She also enjoys discovering new and innovative products to recommend to her friends. She fills her basket with healthy and environmentally friendly products and leaves the store satisfied with her choices.\\
\hline
\end{tabular}
\end{table}

Considering the most common current practice scenarios, the following claims emerge. Consumers unaware of the footprint in their purchasing decision-making process evaluate products just on convenience and personal preferences. Being aware of their environmental impact, consumers could introduce an additional element of discrimination on a par with the previous conditions. Instead, those who have little time to browse products can benefit from the advice of others who also assess the environmental impacts of a product before making a purchase. Finally, it is worth noting that everyone has a smartphone at their disposal, which they use for various purposes.

\begin{figure}[htp]
    \centering
    \subfloat[Maria's behavior.\label{fig:transition_matrix_p}]{%
      \includegraphics[width=0.5\textwidth]{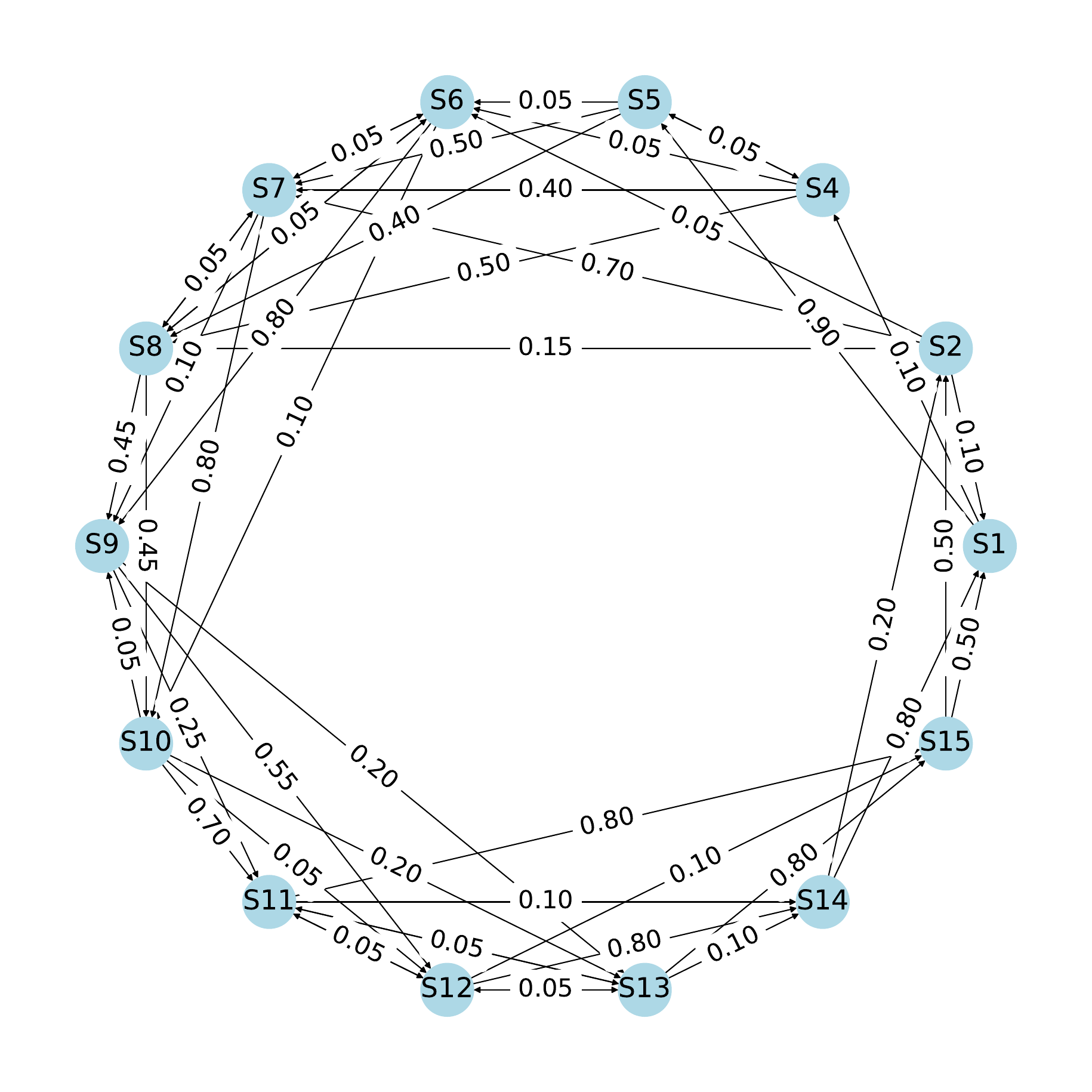}%
    }\hfil
    \subfloat[Olivia's behavior.\label{fig:transition_matrix_q}]{%
      \includegraphics[width=0.5\textwidth]{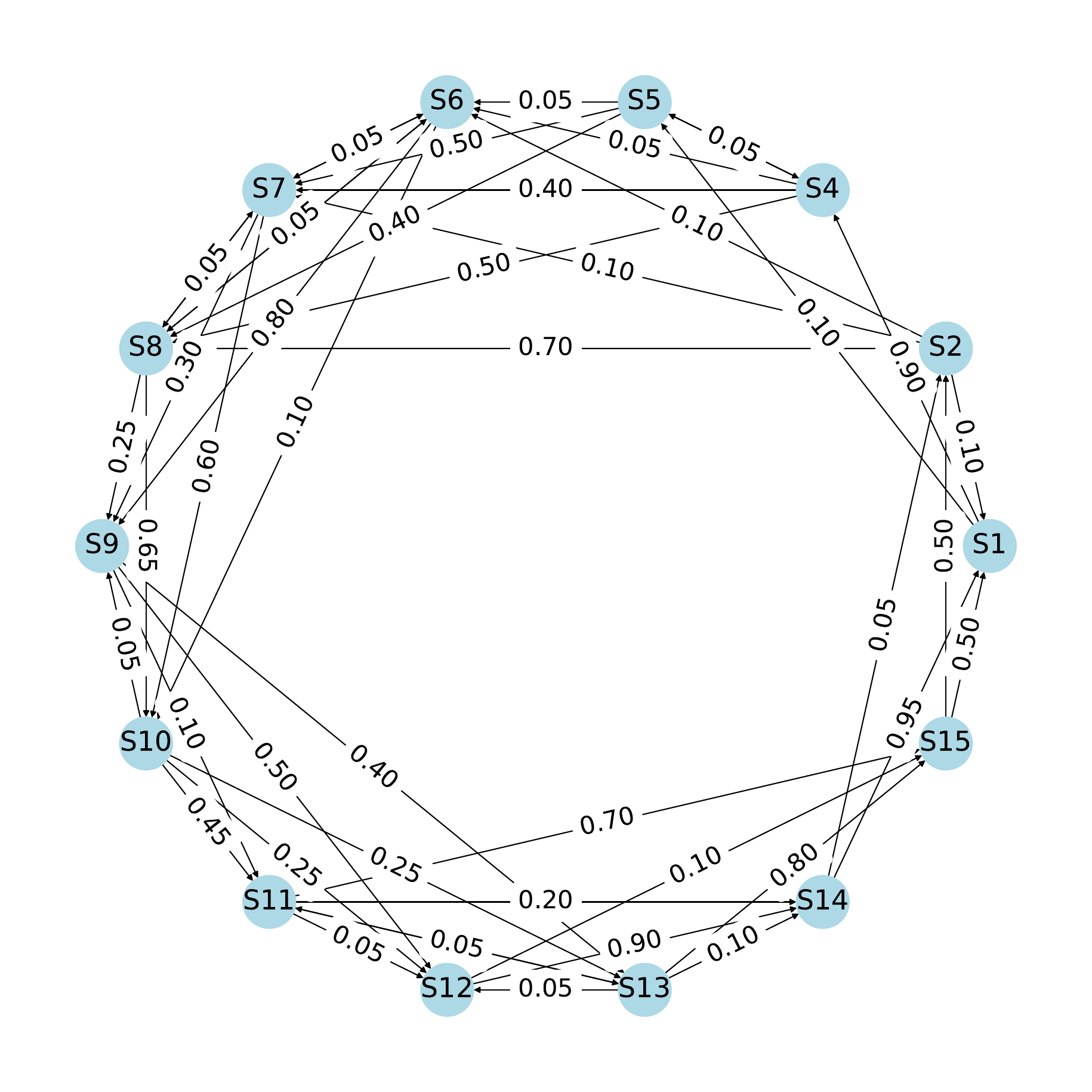}%
    }
    \caption{State transition diagrams of current practices.}
    \label{fig:transition_matrices}
\end{figure}

We can formally describe the behavior of Maria and Olivia using a Markov chain, with the states outlined in Table~\ref{tab:states}. In a Markov chain, the probability of each event depends solely on the immediately preceding event. Each consumer transitions between states with varying probabilities based on their behavior. For a generic transition matrix $M$, we denote the probability of transitioning from state $i$ to state $j$ as $M_{(i)(j)}$. The transition matrices $P$ and $Q$, which represent the behavior of Maria and Olivia in the current scenarios, are depicted as transition diagrams in Fig.~\ref{fig:transition_matrix_p} and Fig.~\ref{fig:transition_matrix_q}, respectively.

\begin{table}[ht!]
    \small
    \renewcommand{\arraystretch}{1.2}
    \caption{States in the Markov Chain}
    \label{tab:states}
    \centering
    \begin{tabularx}{\linewidth}{l c X }
    \toprule
    \textbf{Macro-state} & \textbf{State } & \textbf{Description} \\
    \midrule
    \multirow{2}{*}{Preparation} & S1 & Prepares the shopping list \\
     & S2 & Does not prepare a shopping list \\
    \hline
    \multirow{4}{*}{Support} & S3 & Prepares the shopping list using the application with the proposed interface \\
     & S4 & Prepares the shopping list with a generic checklist \\
     & S5 & Prepares the shopping list with pen and paper \\
    \hline
    \multirow{3}{*}{Influence on purchases} & S6 & Exposure to recommendations of low environmental impact products by other users \\
     & S7 & Habitual purchases \\
     & S8 & Other influences \\
    \hline
    \multirow{2}{*}{Purchase attention} & S9 & Reads product labels \\
     & S10 & Does not read product labels \\
    \hline
    \multirow{3}{*}{Item comparison} & S11 & Compares products based on price \\
     & S12 & Compares products based on environmental impact \\
     & S13 & Compares products based on other characteristics \\
    \hline
    \multirow{4}{*}{Sharing} & S14 & Shares recommendations on low environmental impact products \\
     & S15 & Does not share recommendations on low environmental impact products \\
    \bottomrule
    \end{tabularx}
\end{table}

\subsection{Application requirements}
Scenarios of current practices allowed us to produce the first set of user interface requirements. We describe them below, classifying them into five categories according to Preece et al. \cite{preece2023interaction}.

\paragraph{Functional requirements}
\begin{itemize}
    \item The application allows users to manage a grocery list. This requirement satisfies the general need to store items to be purchased.
    \item The application allows users to cross out a list item by framing a product identification code (e.g., a barcode or QR-code). This requirement allows users to discover directly the environmental impact of the products they purchase.
    \item When a product with a high environmental impact is scanned, the application suggests a similar product with a lower environmental impact. This requirement raises awareness of the need to choose low-impact products.
    \item The application allows users to share information about their purchases with the community. This requirement promotes civic engagement in tackling climate change.
\end{itemize}

\paragraph{Data requirements}
\begin{itemize}
    \item The application must be able to handle large amounts of data efficiently. This requirement allows user data to be collected, processed and shared without slowing down or crashing.
    \item The application must be able to manage data synchronization between users. This requirement allows users to share up-to-date information in real-time.
\end{itemize}

\paragraph{Environmental requirements}
\begin{itemize}
    \item The application must be compatible with the most popular mobile devices. This requirement ensures broad distribution and use by a large segment of users.
    \item The application must be usable in environments with different lighting conditions. This requirement ensures visibility and readability in different environmental conditions, including those with low lighting.
\end{itemize}

\paragraph{User requirements}
\begin{itemize}
    \item Users should have prior experience with utilizing smartphones for basic tasks.
    \item The target user age group predominantly ranges from 18 to 60 years old.
\end{itemize}

\paragraph{Usability requirements}
\begin{itemize}
    \item The application must be sufficiently easy to use by requiring no specific training effort. This requirement ensures that the app is usable by anyone with a minimum effort.
    \item The user interface should provide simple management of user mistakes. This requirement reduces frustration due to unintentional user errors.
    \item The application must be usable by people with different physical abilities. In particular, the application should be accessible to visually impaired users by supporting features such as text-to-speech or operating system font enlargement. This requirement ensures access for all citizens without discrimination.
\end{itemize}

\subsection{Gamification elements}
Gamification involves incorporating game elements into non-game contexts to incentivize the use of a system by tapping into the naturally rewarding aspects of games \cite{deterding2011game}. We introduced gamification elements into the application, such as points, badges, levels and leaderboard, to help motivate and engage users.

A distinction between intrinsic and extrinsic motivation is well-known in the literature \cite{malone1981toward}. Intrinsic motivation is the most authentic and self-directed form of motivation, where individuals engage in an activity for the sheer pleasure and personal interest it arouses because they find the activity intrinsically rewarding. On the other hand, extrinsic motivation is based on external rewards, such as a reward or money or the avoidance of a negative consequence. 

The Self-determination Theory (SDT) \cite{ryan2000self} posits that the intrinsic motivation of individuals is influenced by three fundamental psychological needs: Competence, Relatedness, and Autonomy. Competence involves the acquisition of skills to deal effectively with the external environment, including tasks such as solving challenging problems. Relatedness refers to the significance of social connections, encompassing interactions and competition with others. Lastly, Autonomy represents an intrinsic desire to have control over one's life and to act in accordance with own values.

We mapped the gamified elements of the app and their motivation to the three psychological needs defined in Self-Determination Theory (SDT), as shown in Table~\ref{tab:std_mapping}.

\begin{table}[ht]
\small
\setlength{\tabcolsep}{8pt}
\renewcommand{\arraystretch}{1.05}
\centering
\caption{Gamified features of the application}\label{tab:std_mapping}
\begin{tabular}{p{2cm}p{6.7cm}p{1.7cm}}
\toprule
\textbf{Feature} & \textbf{Rationale} & \textbf{SDT} \\
\midrule
Points, Levels & Users earn points scanning products. Users' awareness levels up based on the points they earn. & Competence \\ 
\hline
Mission & A mission has a specific objective, e.g., identifying five products with a smaller environmental footprint in the soft drinks category. & Competence, Autonomy \\
\hline
Badges & When users complete a specific mission they earn a badge. Badges are displayed on the user's profile. & Competence, Relatedness \\
\hline
Leaderboard, User profile & Users can share their progress and grocery list items with other players. & Relatedness \\
\bottomrule
\end{tabular}
\end{table}

\subsection{Scenarios transformation}
An app that assists users with grocery lists may help consumers make more environmentally conscious choices. Since all consumers are familiar with smartphones for other tasks, the development of a mobile application seems to be an appropriate direction. The idea behind the application has led to a transformation of the scenario reported in Table~\ref{tab:transformed_scenarios}.
\begin{table}[t]
\small
\setlength{\tabcolsep}{10pt}
\renewcommand{\arraystretch}{1.5}
\caption{Activity transformation scenarios}\label{tab:transformed_scenarios}
\begin{tabular}{p{0.95\textwidth}}
\hline
(1) Maria, knowing that she can find suggestions for quickly creating a grocery list, decides to try the application with the proposed interface. Now she can see the environmental footprint of each product as she puts items into her basket. When Maria has to choose between two products of the same price, she chooses the one with the lowest food footprint.\\
\hline
(2) Olivia, a conscientious shopper, consistently ranks at the top of the leaderboard for users who make environmentally conscious purchasing decisions based on the food footprint of each product. Olivia publicly shares the choices that have allowed her to rank at the top of the leaderboard with other users.\\
\hline
\end{tabular}
\end{table}

\begin{figure}[htp]
    \centering
    \subfloat[Maria's behavior.\label{fig:stationary_distribution_p}]{%
      \includegraphics[width=0.5\textwidth]{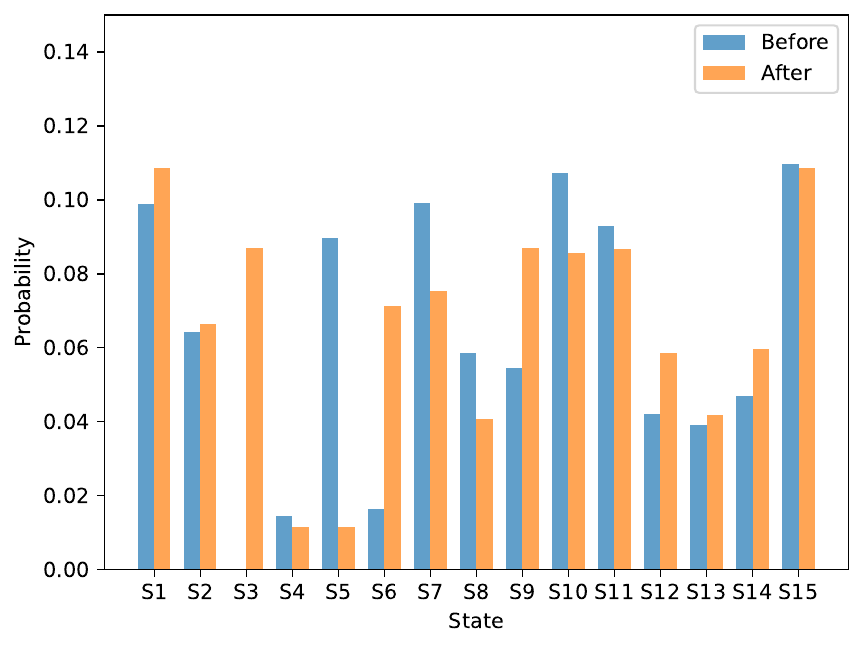}%
    }\hfil
    \subfloat[Olivia's behavior.\label{fig:stationary_distribution_q}]{%
      \includegraphics[width=0.5\textwidth]{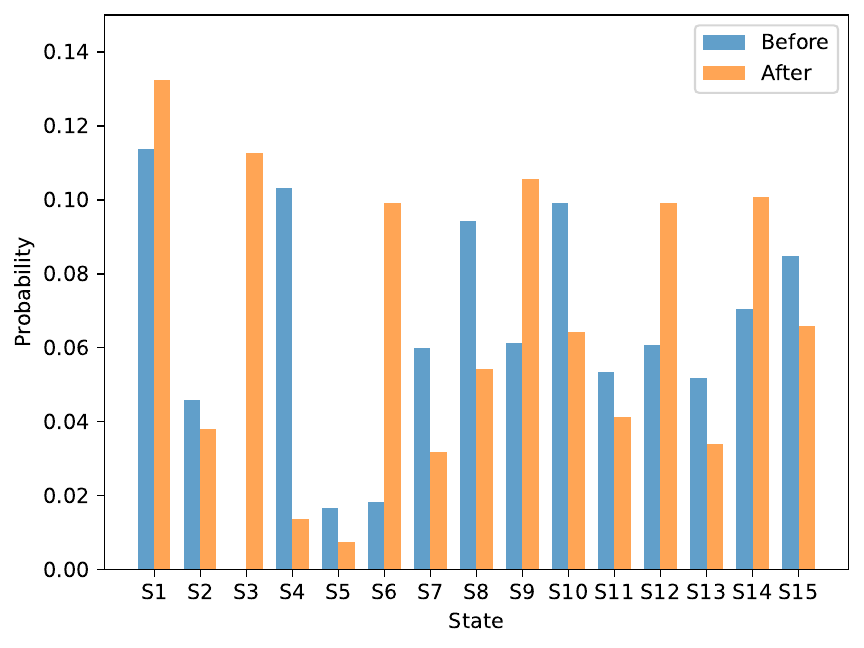}%
    }
    \caption{Stationary distribution of the Markov chain before and after adoption of the application with the proposed interface.}
    \label{fig:stationary_distribution}
\end{figure}

As for the Markov analysis process, the transition diagrams depicted in Fig.~\ref{fig:transition_matrices} undergo an evolution through the inclusion of the state S3, which is described in Table \ref{tab:states}.
Specifically, in Fig.~\ref{fig:transition_matrix_p}, we have updated the following transitions: $P_{(S1)(S3)}=0.80$, $P_{(S1)(S5)}=0.10$, $P_{(S3)(S6)}=0.70$, $P_{(S3)(S7)}=0.15$, and $P_{(S3)(S8)}=0.15$. In Fig.~\ref{fig:transition_matrix_q}, the updates encompass the transitions $Q_{(S1)(S3)}=0.85$, $Q_{(S1)(S4)}=0.10$, $Q_{(S1)(S5)}=0.05$, $Q_{(S3)(S6)}=0.80$, $Q_{(S3)(S7)}=0.10$, $Q_{(S3)(S8)}=0.10$, $Q_{(S9)(S12)}=0.75$, and $Q_{(S9)(S13)}=0.15$.

By calculating the stationary distribution through numerical experiments, we can analyze the changes in user behavior resulting from the adoption of the application. From the graphs in Fig.~\ref{fig:stationary_distribution}, it is evident that, in general, the likelihood of being in states S6, S9, S12, and S14 has increased. 
Specifically, Maria tends to be more careful when shopping. Reading labels and comparing products makes her aware of the environmental impact of her purchases.
Olivia, on the other hand, now has a platform where she can easily share her green choices. Olivia knows that, in her own small way, her contributions can positively influence the choices of other users, including Maria.

Based on the described activity scenarios, we identified the most significant design requirements for the application.
\begin{itemize}
    \item The presentation of the shopping list must be simple and immediate, leveraging the communicative power of images and the language of colors. This allows the small mobile device to provide users with composed and complex information.
    \item The interface must provide users with a reduced number of tabs for navigating between screens. Users use this application frequently and require rapid usage.
    \item Checking products at the time of purchase must occur quickly and immediately, as should the display of similar products.
\end{itemize}

\subsection{Designing the Mobile User Interface}
Based on the scenarios, claims, and application requirements, we have developed a set of interface requirements to guide the design of our application. 
\begin{itemize}
    \item \textbf{Interface that minimizes text input}. As the user starts typing, product suggestions containing the substring of the input text based on previous shopping lists are displayed. Frequent and latest products are also suggested.
    \item \textbf{Product list with images}. Users can quickly identify the product they are looking for regardless of their cultural background.
    \item \textbf{Interface that limits the number of interactions}. Only the necessary elements for completing a task are shown, while everything else is hidden.
    \item \textbf{Identification code scanner}. Users can quickly select purchased items by scanning the barcode or QR code available on its packaging.
    \item  \textbf{Product footprint label}. After the user scans the identification code of a product, a star rating system \cite{leach2016environmental} with the footprint is shown.
\end{itemize}

In terms of the application's information architecture, when users enter the application, they can create a new shopping list, and suggestions are made based on their previous lists to help them create a new one. Suggestions may include low-carbon alternatives based on the purchases of other users.

\begin{figure}
    \includegraphics[width=0.9\textwidth]{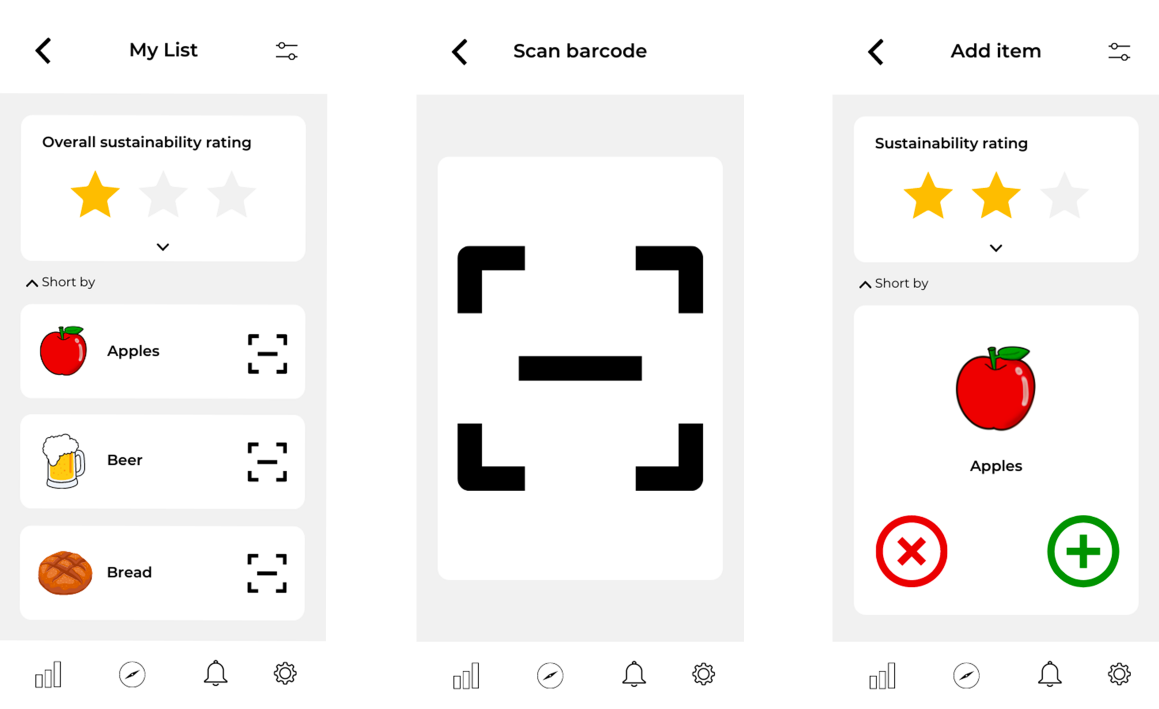}
    \caption{Application User Interface prototype} \label{fig:ui}
\end{figure}

Once a grocery list is created, users can view it and mark the products added to their basket by scanning the relevant identification code. As each product is scanned, its food product footprint is displayed, increasing the user's awareness of its impact on the environment. Fig.~\ref{fig:ui} shows the application screens relating to a shopping list, the scanning of a product, and the crossing out of a product to be added to the basket.

\section{Discussion and Conclusions}
\label{sec:conclusions}
The development of mobile applications is an effective way to increase user awareness and engagement in activities of common interest. One of the main challenges is to create a user-friendly and usable interface for all types of users.

The design requirement that guided our work is to make consumers aware of their impact on climate change directly and tangibly. The artifact is a mobile application that enables users to generate a shopping list while assigning an estimated food product footprint and allows users to be aware of other products similar to the ones they use but with a lower footprint. 

The application prototype has been designed using user-centered design principles. The scenario analysis helped to identify the different actors involved and how they interact with the products during the purchase process.

The next step in this research will involve conducting a field usability test with customers in a supermarket to demonstrate the effectiveness of the initial solution. Further evaluation and testing are necessary to determine the effectiveness of the artifact in achieving its intended goals and to identify any areas for improvement.

\bibliographystyle{splncs04}
\bibliography{references}

\end{document}